\documentclass[prd,aps,showpacs,epsf,floats]{revtex4}
\usepackage{amsfonts}
\usepackage{amsmath}

\input{tcilatex}

\begin{document}

\title{Information-Geometric Indicators of Chaos in Gaussian Models on
Statistical Manifolds of Negative Ricci Curvature}
\author{Carlo Cafaro${}$}
\email{carlocafaro2000@yahoo.it}
\affiliation{Department of Physics, State University of New York at Albany-SUNY,1400
Washington Avenue, Albany, NY 12222, USA}

\begin{abstract}
A new information-geometric approach to chaotic dynamics on curved
statistical manifolds based on Entropic Dynamics (ED) is proposed. It is
shown that the hyperbolicity of a non-maximally symmetric $6N$-dimensional
statistical manifold $\mathcal{M}_{s}$ underlying an ED Gaussian model
describing an arbitrary system of $3N$ degrees of freedom leads to linear
information-geometric entropy growth and to exponential divergence of the
Jacobi vector field intensity, quantum and classical features of chaos
respectively.
\end{abstract}

\maketitle


\section{Introduction}

The unification of classical theory of gravity with quantum theories of
electromagnetic, weak and strong forces is one of the major problems in
modern physics. Entropic Dynamics (ED) \cite{catichaED}, namely the
combination of principles of inductive inference (Maximum relative Entropy
Methods, ME methods) \cite{caticha(REII), caticha-giffin, caticha-preuss}
and methods of Information Geometry (Riemannian geometry applied to
probability theory, IG) \cite{amari}, is a theoretical framework constructed
on statistical manifolds and it is developed to investigate the possibility
that laws of physics, either classical or quantum, might reflect laws of
inference rather than laws of nature. Examples of dynamics that can be
deduced from principles of probable inference are not absent in physics. The
theory of thermodynamics \cite{jaynes} and to a \ certain degree, quantum
mechanics \cite{caticha(PLA)}, are examples of fundamental physical theories
that could be derived from general principles of inference. In constructing
an ED-model, the first step is to identify the appropriate variables
(relevant information) describing the system and thus the corresponding
space of macrostates. This is the most delicate step because there is no
systematic way to search for the right macro variables; it is a matter of
intuition, trial and error. Once the information and normalization
constraints are identified, using ME methods, the probability distribution
characterizing the system can be computed. Finally, using IG methods, a
Fisher-Rao information metric \cite{fisher, rao} can be assigned to the
space of macrostates of the system. Given the Fisher-Rao information metric,
the geometric structure of the manifold underlying the ED can be studied in
detail: metric tensor, Christoffel connection coefficients, Ricci and
Riemann statistical curvature tensors, sectional and Ricci scalar
curvatures, Jacobi and Killing fields can be calculated. ME methods are
inductive inference tools. They are used for updating from a prior to a
posterior distribution when new information in the form of constraints
becomes available. Basically, information is processed using ME methods in
the framework of Information Geometry. The ED model follows from an
assumption about what information is relevant to predict the evolution of
the system. In this work, we focus only on reversible aspects of the ED
model. In this case, given a known initial macrostate and that the system
evolves to a final known macrostate, we investigate the possible
trajectories of the system. Given two probability distributions, a notion of
"distance" between them is provided by IG.

In this paper, our objective is to report some relevant results obtained in
the realm of chaos theory (hypersensitivity to initial conditions) using the
ED formalism. It is known there is not a well defined unifying
characterization of chaos in classical and quantum physics \cite{caves,
scott}. In the Riemannian Geometric Approach \cite{casetti, di bari} to
classical chaos, the search for a link between the Jacobi field intensity
and the Ricci (sectional) curvature of the dynamical manifold is under
investigation \cite{kawabe}. In the Zurek-Paz criterion of quantum chaos 
\cite{zurek}, instead, the search for a potential link between the linearity
of the entropy growth and the curvature of the dynamical manifold underlying
chaotic systems is still open. In our information-geometric approach, it is
shown that these three indicators of chaos (Curvature-Jacobi Field
Intensity-Entropy) are linked: the hyperbolicity of a $6N$-dimensional
statistical manifold $\mathcal{M}_{s}$ underlying an ED Gaussian model leads
to linear information-geometric entropy growth and to exponential divergence
of the Jacobi vector field intensity, quantum and classical features of
chaos respectively. We assume that the arbitrary (physical-biological)
system under investigation has $3N$ degrees of freedom $\left\{
x_{a}^{\left( \alpha \right) }\right\} _{a=1\text{, }2\text{, }3}^{\alpha =1%
\text{,.., }N}$, each one described by two pieces of relevant information,
its expectation value $\left\langle x_{a}^{\left( \alpha \right)
}\right\rangle $ and variance $\Delta x_{a}^{\left( \alpha \right) }\overset{%
\text{def}}{=}\sqrt{\left\langle \left( x_{a}^{\left( \alpha \right)
}-\left\langle x_{a}^{\left( \alpha \right) }\right\rangle \right)
^{2}\right\rangle }$. This leads to consider an ED model on a $6N$%
-dimensional statistical manifold $\mathcal{M}_{s}$. First, we show that $%
\mathcal{M}_{s}$\ has a constant negative Ricci curvature proportional to
the number of degrees of freedom of the system, $\mathcal{R}_{\mathcal{M}%
_{s}}=-3N$. Second, we suggest the information-geometric analog of the
Zurek-Paz quantum chaos criterion. It is shown that the system explores
statistical volume elements on $\mathcal{M}_{s}$\ at an exponential rate. We
define an information-geometric entropy (IGE) of the system, $\mathcal{S}_{%
\mathcal{M}_{s}}$. We show that $\mathcal{S}_{\mathcal{M}_{s}}$\ increases
linearly in time (statistical evolution parameter) and it is proportional to
the number of degrees of freedom of the system and to the
information-geometric analogue of the Lyapunov exponents \cite{ruelle}.
Finally, we show that the geodesics on the manifold $\mathcal{M}_{s}$\ are
described by hyperbolic trajectories. Using the Jacobi-Levi-Civita (JLC)
equation for geodesic spread, it is shown that the intensity of the Jacobi
vector field intensity $J_{\mathcal{M}_{S}}$\ diverges exponentially
(standard feature of classical chaos) and it is proportional to the number
of degrees of freedom of the system. In conclusion, the Ricci scalar
curvature $\mathcal{R}_{\mathcal{M}_{s}}$, the information-geometric entropy 
$\mathcal{S}_{\mathcal{M}_{s}}$\ and the Jacobi vector field intensity $J_{%
\mathcal{M}_{S}}$\ are proportional to the number of Gaussian-distributed
microstates of the system. The relevance of this proportionality will be
discussed in some detail in section III of this article.

The layout of this paper is as follows. In section II, we describe the ED
Gaussian model being studied. In section III, we introduce the
information-geometric indicators of chaos for our theoretical model.
Finally, in section IV we present our final remarks.

\section{Entropic Dynamical Gaussian Model}

We consider an ED model whose microstates span a $3N$-dimensional space
labelled by the variables $\left\{ \vec{X}\right\} =\left\{ \vec{x}^{\left(
1\right) }\text{, }\vec{x}^{\left( 2\right) }\text{,...., }\vec{x}^{\left(
N\right) }\right\} $ with $\vec{x}^{\left( \alpha \right) }\equiv \left(
x_{1}^{\left( \alpha \right) }\text{, }x_{2}^{\left( \alpha \right) }\text{, 
}x_{3}^{\left( \alpha \right) }\right) $, $\alpha =1$,...., $N$ and $%
x_{a}^{\left( \alpha \right) }\in 
\mathbb{R}
$ with $a=1$, $2$, $3$ . We assume the only testable information pertaining
to the quantities $x_{a}^{\left( \alpha \right) }$ consists of the
expectation values $\left\langle x_{a}^{\left( \alpha \right) }\right\rangle 
$ and the variance $\Delta x_{a}^{\left( \alpha \right) }$. The set of these
expected values define the $6N$-dimensional space of macrostates of the
system. A measure of distinguishability among the macrostates of the ED
model is achieved by assigning a probability distribution $P\left( \vec{X}%
\left\vert \vec{\Theta}\right. \right) $ to each $6N$-dimensional macrostate 
$\vec{\Theta}\overset{\text{def}}{=}\left\{ \left( ^{\left( 1\right) }\theta
_{a}^{\left( \alpha \right) }\text{,}^{\left( 2\right) }\theta _{a}^{\left(
\alpha \right) }\right) \right\} _{3N\text{-pairs}}$ $=\left\{ \left(
\left\langle x_{a}^{\left( \alpha \right) }\right\rangle \text{, }\Delta
x_{a}^{\left( \alpha \right) }\right) \right\} _{3N\text{-pairs}}$with $%
\alpha =1$, $2$,...., $N$ and $a=1$, $2$, $3$. The process of assigning a
probability distribution to each state provides $\mathcal{M}_{S}$ with a
metric structure. Specifically, the Fisher-Rao information metric defined in
(\ref{FRM}) is a measure of distinguishability among macrostates. It assigns
an IG to the space of states. Consider an arbitrary physical system evolving
over a $3N$-dimensional space.\ The variables $\left\{ \vec{X}\right\} 
\overset{\text{def}}{=}\left\{ \vec{x}^{\left( 1\right) }\text{, }\vec{x}%
^{\left( 2\right) }\text{,...., }\vec{x}^{\left( N\right) }\right\} $ label
the $3N$-dimensional space of microstates of the system. We assume that all
information relevant to the dynamical evolution of the system is contained
in the probability distributions. For this reason, no other information is
required. Each macrostate may be thought as a point of a $6N$-dimensional
statistical manifold with coordinates given by the numerical values of the
expectations $^{\left( 1\right) }\theta _{a}^{\left( \alpha \right) }$ and $%
^{\left( 2\right) }\theta _{a}^{\left( \alpha \right) }$. The available
information can be written in the form of the following $6N$ information
constraint equations,%
\begin{equation}
\begin{array}{c}
\left\langle x_{a}^{\left( \alpha \right) }\right\rangle
=\dint\limits_{-\infty }^{+\infty }dx_{a}^{\left( \alpha \right)
}x_{a}^{\left( \alpha \right) }P_{a}^{\left( \alpha \right) }\left(
x_{a}^{\left( \alpha \right) }\left\vert ^{\left( 1\right) }\theta
_{a}^{\left( \alpha \right) }\text{,}^{\left( 2\right) }\theta _{a}^{\left(
\alpha \right) }\right. \right)  \\ 
\\ 
\Delta x_{a}^{\left( \alpha \right) }=\left[ \dint\limits_{-\infty
}^{+\infty }dx_{a}^{\left( \alpha \right) }\left( x_{a}^{\left( \alpha
\right) }-\left\langle x_{a}^{\left( \alpha \right) }\right\rangle \right)
^{2}P_{a}^{\left( \alpha \right) }\left( x_{a}^{\left( \alpha \right)
}\left\vert ^{\left( 1\right) }\theta _{a}^{\left( \alpha \right) }\text{,}%
^{\left( 2\right) }\theta _{a}^{\left( \alpha \right) }\right. \right) %
\right] ^{\frac{1}{2}}\text{.}%
\end{array}
\label{C1}
\end{equation}%
The probability distributions $P_{a}^{\left( \alpha \right) }$ in (\ref{C1})
are constrained by the conditions of normalization,%
\begin{equation}
\dint\limits_{-\infty }^{+\infty }dx_{a}^{\left( \alpha \right)
}P_{a}^{\left( \alpha \right) }\left( x_{a}^{\left( \alpha \right)
}\left\vert ^{\left( 1\right) }\theta _{a}^{\left( \alpha \right) }\text{,}%
^{\left( 2\right) }\theta _{a}^{\left( \alpha \right) }\right. \right) =1%
\text{.}  \label{C2}
\end{equation}%
Information theory identifies the Gaussian distribution as the maximum
entropy distribution if only the expectation value and the variance are
known \cite{jaynes2}. ME methods \cite{caticha(REII), caticha-giffin,
caticha-preuss} allow us to associate a probability distribution $P\left( 
\vec{X}\left\vert \vec{\Theta}\right. \right) $ to each point in the space
of states $\vec{\Theta}$. The distribution that best reflects the
information contained in the prior distribution $m\left( \vec{X}\right) $
updated by the information $\left( \left\langle x_{a}^{\left( \alpha \right)
}\right\rangle ,\Delta x_{a}^{\left( \alpha \right) }\right) $ is obtained
by maximizing the relative entropy 
\begin{equation}
S\left( \vec{\Theta}\right) =-\int d^{3N}\vec{X}P\left( \vec{X}\left\vert 
\vec{\Theta}\right. \right) \log \left( \frac{P\left( \vec{X}\left\vert \vec{%
\Theta}\right. \right) }{m\left( \vec{X}\right) }\right) \text{,}  \label{RE}
\end{equation}%
where $m(\vec{X})$ is the prior probability distribution. As a working
hypothesis, the prior $m\left( \vec{X}\right) $ is set to be uniform since
we assume the lack of prior available information about the system
(postulate of equal \textit{a priori} probabilities). Upon maximizing (\ref%
{RE}), given the constraints (\ref{C1}) and (\ref{C2}), we obtain%
\begin{equation}
P\left( \vec{X}\left\vert \vec{\Theta}\right. \right) =\dprod\limits_{\alpha
=1}^{N}\dprod\limits_{a=1}^{3}P_{a}^{\left( \alpha \right) }\left(
x_{a}^{\left( \alpha \right) }\left\vert \mu _{a}^{\left( \alpha \right) }%
\text{, }\sigma _{a}^{\left( \alpha \right) }\right. \right)   \label{PDG}
\end{equation}%
where%
\begin{equation}
P_{a}^{\left( \alpha \right) }\left( x_{a}^{\left( \alpha \right)
}\left\vert \mu _{a}^{\left( \alpha \right) }\text{, }\sigma _{a}^{\left(
\alpha \right) }\right. \right) =\left( 2\pi \left[ \sigma _{a}^{\left(
\alpha \right) }\right] ^{2}\right) ^{-\frac{1}{2}}\exp \left[ -\frac{\left(
x_{a}^{\left( \alpha \right) }-\mu _{a}^{\left( \alpha \right) }\right) ^{2}%
}{2\left( \sigma _{a}^{\left( \alpha \right) }\right) ^{2}}\right] 
\end{equation}%
and,\textbf{\ }in standard notation for Gaussians, $^{\left( 1\right)
}\theta _{a}^{\left( \alpha \right) }\overset{\text{def}}{=}\left\langle
x_{a}^{\left( \alpha \right) }\right\rangle \equiv \mu _{a}^{\left( \alpha
\right) }$, $^{\left( 2\right) }\theta _{a}^{\left( \alpha \right) }\overset{%
\text{def}}{=}\Delta x_{a}^{\left( \alpha \right) }\equiv \sigma
_{a}^{\left( \alpha \right) }$. The probability distribution (\ref{PDG})
encodes the available information concerning the system. Note that we have
assumed uncoupled constraints among microvariables $x_{a}^{\left( \alpha
\right) }$. In other words, we assumed that information about correlations
between the microvariables need not to be tracked. This assumption leads to
the simplified product rule (\ref{PDG}). However, coupled constraints would
lead to a generalized product rule in (\ref{PDG}) and to a metric tensor (%
\ref{FRM}) with non-trivial off-diagonal elements (covariance terms).
Correlation terms may be fictitious. They may arise for instance from
coordinate transformations. On the other hand, correlations may arise from
external fields in which the system is immersed. In such situations,
correlations among $x_{a}^{\left( \alpha \right) }$ effectively describe
interaction between the microvariables and the external fields. Such
generalizations would require more delicate analysis.

We cannot determine the evolution of microstates of the system since the
available information is insufficient. Not only is the information available
insufficient but we also do not know the equation of motion. In fact there
is no standard "equation of motion".\ Instead we can ask: how close are the
two total distributions with parameters $(\mu _{a}^{\left( \alpha \right) }$%
, $\sigma _{a}^{\left( \alpha \right) })$ and $(\mu _{a}^{\left( \alpha
\right) }+d\mu _{a}^{\left( \alpha \right) }$, $\sigma _{a}^{\left( \alpha
\right) }+d\sigma _{a}^{\left( \alpha \right) })$? Once the states of the
system have been defined, the next step concerns the problem of quantifying
the notion of change from the macrostate $\vec{\Theta}$ to the macrostate $%
\vec{\Theta}+d\vec{\Theta}$. A convenient measure of change is distance. The
measure we seek is given by the dimensionless "distance" $ds$ between $%
P\left( \vec{X}\left\vert \vec{\Theta}\right. \right) $ and $P\left( \vec{X}%
\left\vert \vec{\Theta}+d\vec{\Theta}\right. \right) $,%
\begin{equation}
ds^{2}=g_{\mu \nu }d\Theta ^{\mu }d\Theta ^{\nu }  \label{LE}
\end{equation}%
where%
\begin{equation}
g_{\mu \nu }=\int d\vec{X}P\left( \vec{X}\left\vert \vec{\Theta}\right.
\right) \frac{\partial \log P\left( \vec{X}\left\vert \vec{\Theta}\right.
\right) }{\partial \Theta ^{\mu }}\frac{\partial \log P\left( \vec{X}%
\left\vert \vec{\Theta}\right. \right) }{\partial \Theta ^{\nu }}
\label{FRM}
\end{equation}%
is the Fisher-Rao metric \cite{fisher, rao}. Substituting (\ref{PDG}) into (%
\ref{FRM}), the metric $g_{\mu \nu }$ on $\mathcal{M}_{s}$ becomes a $%
6N\times 6N$ matrix $M$ made up of $3N$ blocks $M_{2\times 2}$ with
dimension $2\times 2$ given by,%
\begin{equation}
M_{2\times 2}=\left( 
\begin{array}{cc}
\left( \sigma _{a}^{\left( \alpha \right) }\right) ^{-2} & 0 \\ 
0 & 2\times \left( \sigma _{a}^{\left( \alpha \right) }\right) ^{-2}%
\end{array}%
\right) 
\end{equation}%
with $\alpha =1$, $2$,...., $N$ and $a=1$, $2$, $3$. From (\ref{FRM}), the
"length" element (\ref{LE}) reads,%
\begin{equation}
ds^{2}=\dsum\limits_{\alpha =1}^{N}\dsum\limits_{a=1}^{3}\left[ \frac{1}{%
\left( \sigma _{a}^{\left( \alpha \right) }\right) ^{2}}d\mu _{a}^{\left(
\alpha \right) 2}+\frac{2}{\left( \sigma _{a}^{\left( \alpha \right)
}\right) ^{2}}d\sigma _{a}^{\left( \alpha \right) 2}\right] \text{.}
\label{LES}
\end{equation}%
We bring attention to the fact that the metric structure of $\mathcal{M}_{s}$
is an emergent (not fundamental) structure. It arises only after assigning a
probability distribution $P\left( \vec{X}\left\vert \vec{\Theta}\right.
\right) $ to each state $\vec{\Theta}$.

\section{Information-Geometric Indicators of Chaos}

In this section, we introduce the relevant indicators of chaoticity within
our theoretical formalism. They are the Ricci scalar curvature $\mathcal{R}_{%
\mathcal{M}_{s}}$\ (or, more correctly, the sectional curvature $\mathcal{K}%
_{\mathcal{M}_{S}}$ \cite{MTW}), the Jacobi vector field intensity $J_{%
\mathcal{M}_{S}}$\ and the IGE $\mathcal{S}_{\mathcal{M}_{s}}$.

\subsection{Ricci Scalar Curvature}

Given the Fisher-Rao information metric, we use standard differential
geometry methods applied to the space of probability distributions to
characterize the geometric properties of $\mathcal{M}_{s}$. Recall that the
Ricci scalar curvature $\mathcal{R}$ is given by,%
\begin{equation}
\mathcal{R}=g^{\mu \nu }R_{\mu \nu }\text{,}
\end{equation}%
where $g^{\mu \nu }g_{\nu \rho }=\delta _{\rho }^{\mu }$ so that $g^{\mu \nu
}=\left( g_{\mu \nu }\right) ^{-1}$. The Ricci tensor $R_{\mu \nu }$ is
given by,%
\begin{equation}
R_{\mu \nu }=\partial _{\varepsilon }\Gamma _{\mu \nu }^{\varepsilon
}-\partial _{\nu }\Gamma _{\mu \varepsilon }^{\varepsilon }+\Gamma _{\mu \nu
}^{\varepsilon }\Gamma _{\varepsilon \eta }^{\eta }-\Gamma _{\mu \varepsilon
}^{\eta }\Gamma _{\nu \eta }^{\varepsilon }\text{.}
\end{equation}%
The Christoffel symbols $\Gamma _{\mu \nu }^{\rho }$ appearing in the Ricci
tensor are defined in the standard way, 
\begin{equation}
\Gamma _{\mu \nu }^{\rho }=\frac{1}{2}g^{\rho \varepsilon }\left( \partial
_{\mu }g_{\varepsilon \nu }+\partial _{\nu }g_{\mu \varepsilon }-\partial
_{\varepsilon }g_{\mu \nu }\right) .
\end{equation}%
Using (\ref{LES}) and the definitions given above, we can show that the
Ricci scalar curvature becomes%
\begin{equation}
\mathcal{R}_{\mathcal{M}_{s}}=-3N<0\text{.}  \label{RSC}
\end{equation}%
From (\ref{RSC}) we conclude that $\mathcal{M}_{s}$ is a $6N$-dimensional
statistical manifold of constant negative Ricci scalar curvature. A detailed
analysis on the calculation of Christoffel connection coefficients using the
ED formalism can be found in \cite{cafaro(06), cafaro(PhysicaD)}.
Furthermore, it can be shown that $\mathcal{M}_{s}$ is not a pseudosphere
(maximally symmetric manifold) since its sectional curvature is not
constant. As a final remark, we emphasize that the negativity of the Ricci
scalar\textbf{\ }$\mathcal{R}_{\mathcal{M}_{S}}$\textbf{\ }implies the
existence of expanding directions in the configuration space manifold\textbf{%
\ }$\mathcal{M}_{s}$. Indeed, since\textbf{\ }$\mathcal{R}_{\mathcal{M}_{S}}$%
\textbf{\ }is the sum of all sectional curvatures of planes spanned by pairs
of orthonormal basis elements \cite{MTW}, the negativity of the Ricci scalar
is only a sufficient (not necessary) condition for local instability of
geodesic flow. For this reason, the negativity of the scalar provides a
strong criterion of local instability.

\subsection{Information-Geometrodynamical Entropy}

At this point, we study the trajectories of the system on $\mathcal{M}_{s}$.
We emphasize ED can be derived from a standard principle of least action
(Maupertuis- Euler-Lagrange-Jacobi-type) \cite{catichaED, arnold}. The main
differences are that the dynamics being considered here, namely Entropic
Dynamics, is defined on a space of probability distributions $\mathcal{M}_{s}
$, not on an ordinary linear space $V$ and the standard coordinates $q_{\mu }
$ of the system are replaced by statistical macrovariables $\Theta ^{\mu }$.
The geodesic equations for the macrovariables of the Gaussian ED model are
given by \cite{de felice},%
\begin{equation}
\frac{d^{2}\Theta ^{\mu }}{d\tau ^{2}}+\Gamma _{\nu \rho }^{\mu }\frac{%
d\Theta ^{\nu }}{d\tau }\frac{d\Theta ^{\rho }}{d\tau }=0  \label{GE}
\end{equation}%
with $\mu =1$, $2$,...,$6N$. Observe that the geodesic equations are\textit{%
\ nonlinear}, second order coupled ordinary differential equations. These
equations describe a dynamics that is reversible and their solution is the
trajectory between an initial and a final macrostate. The trajectory can be
equally well traversed in both directions. We seek the explicit form of (\ref%
{GE}) for the pairs of statistical coordinates $(\mu _{a}^{\left( \alpha
\right) }$, $\sigma _{a}^{\left( \alpha \right) })$. Substituting the
explicit expression of the Christoffel connection coefficients into (\ref{GE}%
), the geodesic equations for the macrovariables $\mu _{a}^{\left( \alpha
\right) }$ and $\sigma _{a}^{\left( \alpha \right) }$ associated to the
microstate $x_{a}^{\left( \alpha \right) }$ become,%
\begin{equation}
\text{ }\frac{d^{2}\mu _{a}^{\left( \alpha \right) }}{d\tau ^{2}}-\frac{2}{%
\sigma _{a}^{\left( \alpha \right) }}\frac{d\mu _{a}^{\left( \alpha \right) }%
}{d\tau }\frac{d\sigma _{a}^{\left( \alpha \right) }}{d\tau }=0\text{, }%
\frac{d^{2}\sigma _{a}^{\left( \alpha \right) }}{d\tau ^{2}}-\frac{1}{\sigma
_{a}^{\left( \alpha \right) }}\left( \frac{d\sigma _{a}^{\left( \alpha
\right) }}{d\tau }\right) ^{2}+\frac{1}{2\sigma _{a}^{\left( \alpha \right) }%
}\left( \frac{d\mu _{a}^{\left( \alpha \right) }}{d\tau }\right) ^{2}=0\text{%
.}
\end{equation}%
with $\alpha =1$, $2$,...., $N$ and $a=1$, $2$, $3$. This is a set of
coupled ordinary differential equations, whose solutions are%
\begin{equation}
\mu _{a}^{\left( \alpha \right) }\left( \tau \right) =\frac{\frac{\left(
B_{a}^{\left( \alpha \right) }\right) ^{2}}{2\beta _{a}^{\left( \alpha
\right) }}}{\exp \left( -2\beta _{a}^{\left( \alpha \right) }\tau \right) +%
\frac{\left( B_{a}^{\left( \alpha \right) }\right) ^{2}}{8\left( \beta
_{a}^{\left( \alpha \right) }\right) ^{2}}}\text{, }\sigma _{a}^{\left(
\alpha \right) }\left( \tau \right) =\frac{B_{a}^{\left( \alpha \right)
}\exp \left( -\beta _{a}^{\left( \alpha \right) }\tau \right) }{\exp \left(
-2\beta _{a}^{\left( \alpha \right) }\tau \right) +\frac{\left(
B_{a}^{\left( \alpha \right) }\right) ^{2}}{8\left( \beta _{a}^{\left(
\alpha \right) }\right) ^{2}}}+C_{a}^{\left( \alpha \right) }\text{.}
\label{T}
\end{equation}%
The quantities $B_{a}^{\left( \alpha \right) }$, $C_{a}^{\left( \alpha
\right) }$, $\beta _{a}^{\left( \alpha \right) }$ are real integration
constants and they can be evaluated once the boundary conditions are
specified. We observe that since every geodesic is well-defined for all
temporal parameters $\tau $, $\mathcal{M}_{s}$ constitutes a geodesically
complete manifold \cite{lee}. It is therefore a natural setting within which
one may consider global\textbf{\ }questions and search for a weak criterion
of chaos \cite{di bari}. Furthermore, since $\left\vert \mu _{a}^{\left(
\alpha \right) }\left( \tau \right) \right\vert <+\infty $ and $\left\vert
\sigma _{a}^{\left( \alpha \right) }\left( \tau \right) \right\vert <+\infty 
$ $\forall \tau \in 
\mathbb{R}
^{+}$, $\forall a=1$, $2$, $3$ and $\forall \alpha =1$,.., $N$, the
parameter space $\left\{ \vec{\Theta}\right\} $ (homeomorphic to $\mathcal{M}%
_{s}$) is compact. The compactness of the configuration space manifold $%
\mathcal{M}_{s}$ assures the folding mechanism of information-dynamical
trajectories (the folding mechanism is a key-feature of true chaos, \cite{di
bari}).

We are interested in investigating the stability of the trajectories of the
ED model considered on $\mathcal{M}_{s}$. It is known \cite{arnold} that the
Riemannian curvature of a manifold is closely connected with the behavior of
the geodesics on it. If the Riemannian curvature of a manifold is negative,
geodesics (initially parallel) rapidly diverge from one another. For the
sake of simplicity, we assume very special initial conditions: $%
B_{a}^{\left( \alpha \right) }\equiv \Lambda $, $\beta _{a}^{\left( \alpha
\right) }\equiv \lambda \in 
\mathbb{R}
^{+}$, $C_{a}^{\left( \alpha \right) }=0$, $\forall \alpha =1$, $2$,...., $N$
and $a=1$, $2$, $3$. However, the conclusion we reach can be generalized to
more arbitrary initial conditions. Recall that $\mathcal{M}_{s}$ is the
space of probability distributions $P\left( \vec{X}\left\vert \vec{\Theta}%
\right. \right) $ labeled by $6N$ statistical parameters $\vec{\Theta}$.
These parameters are the coordinates for the point $P$, and in these
coordinates a volume element $dV_{\mathcal{M}_{s}}$ reads, 
\begin{equation}
dV_{\mathcal{M}_{S}}=\sqrt{g}d^{6N}\vec{\Theta}=\dprod\limits_{\alpha
=1}^{N}\dprod\limits_{a=1}^{3}\frac{\sqrt{2}}{\left( \sigma _{a}^{\left(
\alpha \right) }\right) ^{2}}d\mu _{a}^{\left( \alpha \right) }d\sigma
_{a}^{\left( \alpha \right) }
\end{equation}%
where $g=\left\vert \det \left( g_{\mu \nu }\right) \right\vert $. The
volume of an extended region $\Delta V_{\mathcal{M}_{s}}\left( \tau \text{; }%
\lambda \right) $ of $\mathcal{M}_{s}$ is defined by,%
\begin{equation}
\Delta V_{\mathcal{M}_{s}}\left( \tau \text{; }\lambda \right) \overset{%
\text{def}}{=}\dprod\limits_{\alpha
=1}^{N}\dprod\limits_{a=1}^{3}\dint\limits_{\mu _{a}^{\left( \alpha \right)
}\left( 0\right) }^{\mu _{a}^{\left( \alpha \right) }\left( \tau \right)
}\dint\limits_{\sigma _{a}^{\left( \alpha \right) }\left( 0\right) }^{\sigma
_{a}^{\left( \alpha \right) }\left( \tau \right) }\frac{\sqrt{2}}{\left(
\sigma _{a}^{\left( \alpha \right) }\right) ^{2}}d\mu _{a}^{\left( \alpha
\right) }d\sigma _{a}^{\left( \alpha \right) }  \label{VER}
\end{equation}%
where $\mu _{a}^{\left( \alpha \right) }\left( \tau \right) $ and $\sigma
_{a}^{\left( \alpha \right) }\left( \tau \right) $ are given in\textbf{\ }(%
\ref{T}) and where the scalar $\lambda $ is the chosen quantity to define
the one-parameter family of geodesics $\mathcal{F}_{G_{\mathcal{M}%
_{s}}}\left( \lambda \right) \overset{\text{def}}{=}\left\{ \Theta _{%
\mathcal{M}_{s}}^{\mu }\left( \tau \text{; }\lambda \right) \right\}
_{\lambda \in 
\mathbb{R}
^{+}}^{\mu =1\text{,..,}6N}$. The quantity that encodes relevant information
about the stability of neighboring volume elements is the the average volume 
$\mathcal{V}_{\mathcal{M}_{s}}\left( \tau \text{; }\lambda \right) $ defined
as \cite{cafaro(PhysicaD)}, 
\begin{equation}
\mathcal{V}_{\mathcal{M}_{s}}\left( \tau \text{; }\lambda \right) \equiv
\left\langle \Delta V_{\mathcal{M}_{s}}\left( \tau ^{\prime }\text{; }%
\lambda \right) \right\rangle _{\tau }\overset{\text{def}}{=}\frac{1}{\tau }%
\dint\limits_{0}^{\tau }\Delta V_{\mathcal{M}_{s}}\left( \tau ^{\prime }%
\text{; }\lambda \right) d\tau ^{\prime }\overset{\tau \rightarrow \infty }{%
\approx }e^{3N\lambda \tau }\text{.}  \label{AVE}
\end{equation}%
This asymptotic regime of diffusive evolution in (\ref{AVE}) describes the
exponential increase of average volume elements on $\mathcal{M}_{s}$. The
exponential instability characteristic of chaos forces the system to rapidly
explore large areas (volumes) of the statistical manifolds. From equation (%
\ref{AVE}), we notice that the parameter $\lambda $ characterizes the
exponential growth rate of average statistical volumes $\mathcal{V}_{%
\mathcal{M}_{s}}\left( \tau \text{; }\lambda \right) $ in $\mathcal{M}_{s}$.
This suggests that $\lambda $ may play the same role ordinarily played by
Lyapunov exponents. Indeed, it is interesting to note that this asymptotic
behavior appears also in the conventional description of quantum chaos where
the von Neumann entropy increases linearly at a rate determined by the
Lyapunov exponents. The linear entropy increase as a quantum chaos criterion
was introduced by Zurek and Paz \cite{zurek}. In our information-geometric
approach a relevant variable that can be useful to study the degree of
instability characterizing the ED model is the IGE quantity defined as \cite%
{cafaro(PhysicaD)},%
\begin{equation}
\mathcal{S}_{\mathcal{M}_{s}}\overset{\text{def}}{=}\underset{\tau
\rightarrow \infty }{\lim }\log \mathcal{V}_{\mathcal{M}_{s}}\left( \tau 
\text{; }\lambda \right) \text{.}  \label{IGE}
\end{equation}%
The IGE\ is intended to capture the temporal complexity (chaoticity) of ED\
theoretical models on curved statistical manifolds\textbf{\ }$\mathcal{M}_{s}
$ by considering the asymptotic temporal behaviors of the average
statistical volumes occupied by the evolving macrovariables labelling points
on $\mathcal{M}_{s}$.

Substituting (\ref{VER}) in (\ref{AVE}), equation (\ref{IGE}) becomes,%
\begin{equation}
\mathcal{S}_{\mathcal{M}_{s}}=\underset{\tau \rightarrow \infty }{\lim }\log
\left\{ \frac{1}{\tau }\dint\limits_{0}^{\tau }\left[ \dprod\limits_{\alpha
=1}^{N}\dprod\limits_{a=1}^{3}\dint\limits_{\mu _{a}^{\left( \alpha \right)
}\left( 0\right) }^{\mu _{a}^{\left( \alpha \right) }\left( \tau ^{\prime
}\right) }\dint\limits_{\sigma _{a}^{\left( \alpha \right) }\left( 0\right)
}^{\sigma _{a}^{\left( \alpha \right) }\left( \tau ^{\prime }\right) }\frac{%
\sqrt{2}}{\left( \sigma _{a}^{\left( \alpha \right) }\right) ^{2}}d\mu
_{a}^{\left( \alpha \right) }d\sigma _{a}^{\left( \alpha \right) }\right]
d\tau ^{\prime }\right\} \overset{\tau \rightarrow \infty }{\approx }%
3N\lambda \tau \text{.}  \label{IGE(2)}
\end{equation}%
Before discussing the meaning of (\ref{IGE(2)}), recall that in a rigorous
examination of the entropy approach to the classical-quantum correspondence
problem, Zurek and Paz consider the completely tractable model of an
inverted harmonic oscillator with a potential $V\left( x\right) =-\frac{%
\Omega ^{2}x^{2}}{2}$ coupled to a high temperature (harmonic) bath. In
their case, $\Omega $ is analogous to a Lyapunov exponent in a genuinely
chaotic system. On calculating the rate of change of von Neumann entropy,
they show that \cite{zurek}%
\begin{equation}
\mathcal{S}_{\text{von Neumann}}\left( \tau \right) \overset{t\rightarrow
\infty }{\approx }\Omega \tau 
\end{equation}%
where $\mathcal{S}_{\text{von Neumann}}\left( \tau \right) =-tr\left( \rho
_{r}\left( \tau \right) \log \rho _{r}\left( \tau \right) \right) $ is the
von Neumann entropy of the system \cite{benatti} and $\rho _{r}\left( \tau
\right) $ is the reduced density matrix of the system at time $\tau $. The
quantum entropy production rate is determined by the classical instability
parameter\textbf{\ }$\Omega $. Given that the classical Lyapunov exponent to
which $\Omega $ is analogous is equal to the Kolmogorov-Sinai (KS) entropy
of the system \cite{benatti}, this is indeed a remarkable characterization.
It suggests that after a time, a quantum, classically chaotic system loses
information to the environment at a rate determined entirely by the rate at
which the classical system loses information as a result of its dynamics,
namely, the KS entropy.

In analogy to the Zurek-Paz quantum chaos criterion in its classical
reversible limit \cite{zurek(2)}, we suggest a classical \
information-geometric criterion of linear IGE growth. The entropy-like
quantity\textbf{\ }$\mathcal{S}_{\mathcal{M}_{s}}$ \textbf{i}n (\ref{IGE(2)}%
) is the asymptotic limit of the natural logarithm of the statistical weight 
$\left\langle \Delta V_{\mathcal{M}_{s}}\right\rangle _{\tau }$ defined on $%
\mathcal{M}_{s}$ and it grows linearly in time, a quantum feature of chaos.
\ Indeed, equation (\ref{IGE(2)}) may be considered the
information-geometric analog of the Zurek-Paz chaos criterion. In our
chaotic ED\ Gaussian model, the IGE production rate is determined by the
information-geometric parameter $\lambda $ characterizing the exponential
growth rate of average statistical volumes\textbf{\ }$\mathcal{V}_{\mathcal{M%
}_{s}}\left( \tau \text{; }\lambda \right) $ in $\mathcal{M}_{s}$.

\subsection{Jacobi Field Intensity}

Finally, we consider the behavior of the one-parameter family of neighboring
geodesics $\mathcal{F}_{G_{\mathcal{M}_{s}}}\left( \lambda \right) \overset{%
\text{def}}{=}\left\{ \Theta _{\mathcal{M}_{s}}^{\mu }\left( \tau \text{; }%
\lambda \right) \right\} _{\lambda \in 
\mathbb{R}
^{+}}^{\mu =1\text{,..,}6N}$ where,%
\begin{equation}
\mu _{a}^{\left( \alpha \right) }\left( \tau \text{; }\lambda \right) =\frac{%
\frac{\Lambda ^{2}}{2\lambda }}{\exp \left( -2\lambda \tau \right) +\frac{%
\Lambda ^{2}}{8\lambda ^{2}}}\text{, }\sigma _{a}^{\left( \alpha \right)
}\left( \tau \text{; }\lambda \right) =\frac{\Lambda \exp \left( -\lambda
\tau \right) }{\exp \left( -2\lambda \tau \right) +\frac{\Lambda ^{2}}{%
8\lambda ^{2}}}  \label{TR}
\end{equation}%
with $\alpha =1$, $2$,...., $N$ and $a=1$, $2$, $3$. The relative geodesic
spread on a (non-maximally symmetric) curved manifold as $\mathcal{M}_{s}$
is characterized by the Jacobi-Levi-Civita equation, the natural tool to
tackle dynamical chaos \cite{do carmo, MTW},%
\begin{equation}
\frac{D^{2}\delta \Theta ^{\mu }}{D\tau ^{2}}+R_{\nu \rho \sigma }^{\mu }%
\frac{\partial \Theta ^{\nu }}{\partial \tau }\delta \Theta ^{\rho }\frac{%
\partial \Theta ^{\sigma }}{\partial \tau }=0  \label{JLC}
\end{equation}%
where the covariant derivative $\frac{D^{2}\delta \Theta ^{\mu }}{D\tau ^{2}}
$\textbf{\ }in (\ref{JLC}) is defined as \cite{ohanian},%
\begin{eqnarray}
\frac{D^{2}\delta \Theta ^{\mu }}{D\tau ^{2}} &=&\frac{d^{2}\delta \Theta
^{\mu }}{d\tau ^{2}}+2\Gamma _{\alpha \beta }^{\mu }\frac{d\delta \Theta
^{\alpha }}{d\tau }\frac{d\Theta ^{\beta }}{d\tau }+\Gamma _{\alpha \beta
}^{\mu }\delta \Theta ^{\alpha }\frac{d^{2}\Theta ^{\beta }}{d\tau ^{2}}%
+\Gamma _{\alpha \beta ,\nu }^{\mu }\frac{d\Theta ^{\nu }}{d\tau }\frac{%
d\Theta ^{\beta }}{d\tau }\delta \Theta ^{\alpha }+  \notag \\
&&+\Gamma _{\alpha \beta }^{\mu }\Gamma _{\rho \sigma }^{\alpha }\frac{%
d\Theta ^{\sigma }}{d\tau }\frac{d\Theta ^{\beta }}{d\tau }\delta \Theta
^{\rho }\text{,}
\end{eqnarray}%
and the Jacobi vector field $J^{\mu }$ is given by \cite{de felice}\textbf{,}%
\begin{equation}
J^{\mu }\equiv \delta \Theta ^{\mu }\overset{\text{def}}{=}\delta _{\lambda
}\Theta ^{\mu }=\left( \frac{\partial \Theta ^{\mu }\left( \tau \text{; }%
\lambda \right) }{\partial \lambda }\right) _{\tau }\delta \lambda \text{.}
\label{J}
\end{equation}%
Equation (\ref{JLC}) forms a system of $6N$ coupled ordinary differential
equations \textit{linear} in the components of the deviation vector field (%
\ref{J}) but\textit{\ nonlinear} in derivatives of the metric (\ref{FRM}).
It describes the linearized geodesic flow: the linearization ignores the
relative velocity of the geodesics. When the geodesics are neighboring but
their relative velocity is arbitrary, the corresponding geodesic deviation
equation is the so-called generalized Jacobi equation \cite{chicone}. The
nonlinearity is due to the existence of velocity-dependent terms in the
system. Neighboring geodesics accelerate relative to each other with a rate
directly measured by the curvature tensor $R_{\alpha \beta \gamma \delta }$.
Substituting (\ref{TR}) in (\ref{JLC}) and neglecting the exponentially
decaying terms in $\delta \Theta ^{\mu }$ and its derivatives, integration
of (\ref{JLC}) leads to the following asymptotic expression of the Jacobi
vector field intensity,%
\begin{equation}
J_{\mathcal{M}_{S}}=\left\Vert J\right\Vert =\left( g_{\mu \nu }J^{\mu
}J^{\nu }\right) ^{\frac{1}{2}}\overset{\tau \rightarrow \infty }{\approx }%
3Ne^{\lambda \tau }\text{.}
\end{equation}%
Further details on the derivation of this result are in \cite%
{cafaro(PhysicaD)}. We conclude that the geodesic spread on $\mathcal{M}_{s}$
is described by means of an \textit{exponentially} \textit{divergent} Jacobi
vector field intensity $J_{\mathcal{M}_{s}}$, a \textit{classical} feature
of chaos. In our approach the quantity $\lambda _{J}$,%
\begin{equation}
\lambda _{J}\overset{\text{def}}{=}\underset{\tau \rightarrow \infty }{\lim }%
\left[ \frac{1}{\tau }\log \left( \left\Vert \frac{J_{\mathcal{M}_{S}}\left(
\tau \right) }{J_{\mathcal{M}_{S}}\left( 0\right) }\right\Vert \right) %
\right] 
\end{equation}%
would play the role of the conventional Lyapunov exponents.

In conclusion, the main results of this work are encoded in the following
equations,%
\begin{equation}
\mathcal{R}_{\mathcal{M}_{s}}=-3N\text{, }\mathcal{S}_{\mathcal{M}_{s}}%
\overset{\tau \rightarrow \infty }{\approx }3N\lambda \tau \text{, }J_{%
\mathcal{M}_{S}}\overset{\tau \rightarrow \infty }{\approx }3Ne^{\lambda
\tau }\text{.}
\end{equation}%
The IGE grows linearly as a function of the number of Gaussian-distributed
microstates of the system. This supports the fact that\textbf{\ }$\mathcal{S}%
_{\mathcal{M}_{s}}$\ may be a useful measure of temporal complexity \cite{JP}%
. Furthermore, these three indicators of chaoticity, the Ricci scalar
curvature $\mathcal{R}_{\mathcal{M}_{s}}$, the information-geometric entropy 
$\mathcal{S}_{\mathcal{M}_{s}}$ and the Jacobi vector field intensity $J_{%
\mathcal{M}_{S}}$ are proportional to $3N$, the dimension of the microspace
with microstates $\left\{ \vec{X}\right\} $ underlying our chaotic ED\
Gaussian model. This proportionality leads to the conclusion that there is a
substantial link among these information-geometric measures of chaoticity
since they are all extensive functions of the dimensionality of the
microspace underlying the macroscopic chaotic entropic dynamics, namely%
\begin{equation}
\mathcal{R}_{\mathcal{M}_{s}}\sim \mathcal{S}_{\mathcal{M}_{s}}\sim J_{%
\mathcal{M}_{S}}\text{.}  \label{final}
\end{equation}%
Equation (\ref{final})\ represents the fundamental result of this work:
curvature, information-geometric entropy and Jacobi field intensity are
linked within our formalism. We are aware that our findings are reliable in
the restrictive assumption of Gaussianity. However, we believe that with
some additional technical machinery, more general conclusions can be
achieved and this connection among indicators of chaoticity may be
strengthened.

\section{Conclusions}

In this paper, a Gaussian ED statistical model has been constructed on a $6N$%
-dimensional statistical manifold $\mathcal{M}_{s}$. The macrocoordinates on
the manifold are represented by the expectation values of microvariables
associated with Gaussian distributions. The geometric structure of $\mathcal{%
M}_{s}$ was studied in detail. It was shown that $\mathcal{M}_{s}$ is a
curved manifold of constant negative Ricci curvature $-3N$ . The geodesics
of the ED model are hyperbolic curves on $\mathcal{M}_{s}$. A study of the
stability of geodesics on $\mathcal{M}_{s}$ was presented. The notion of
statistical volume elements was introduced to investigate the asymptotic
behavior of a one-parameter family of neighboring volumes $\mathcal{F}_{V_{%
\mathcal{M}_{s}}}\left( \lambda \right) \overset{\text{def}}{=}\left\{ V_{%
\mathcal{M}_{s}}\left( \tau \text{; }\lambda \right) \right\} _{\lambda \in 
\mathbb{R}
^{+}}$. An information-geometric analog of the Zurek-Paz chaos criterion was
suggested. It was shown that the behavior of geodesics is characterized by
exponential instability that leads to chaotic scenarios on the curved
statistical manifold. These conclusions are supported by a study based on
the geodesic deviation equations and on the asymptotic behavior of the
Jacobi vector field intensity $J_{\mathcal{M}_{s}}$ on $\mathcal{M}_{s}$. A
Lyapunov exponent analog similar to that appearing in the Riemannian
geometric approach to chaos was suggested as an indicator of chaoticity. On
the basis of our analysis a relationship among an entropy-like quantity,
chaoticity and curvature is proposed, suggesting to interpret the
statistical curvature as a measure of the entropic dynamical chaoticity. We
think this is a relevant result since a rigorous relation among curvature,
Lyapunov exponents and Kolmogorov-Sinai entropy is still under investigation
and since there does not exist a well defined unifying characterization of
chaos in classical and quantum physics due to fundamental differences
between the two theories. Finally we remark that based on the results
obtained from the chosen ED model, it is not unreasonable to think that
should the correct variables describing the true degrees of freedom of a
physical system be identified, perhaps deeper insights into the foundations
of models of physics and reasoning (and their relationship to each other)
may be uncovered.

\begin{acknowledgments}
The author is grateful to Saleem Ali and Adom Giffin for very useful \
comments. Special thanks go to Ariel Caticha for clarifying explanations on
"Entropic Dynamics" and for his constant support and advice during this
work. Finally, sincere thanks go to the two anonymous Referees for very
helpful suggestions.
\end{acknowledgments}

\end{document}